\documentclass[12pt,preprint]{aastex} 

\begin{document}

\title{
Star Formation in the Bok Globule CB~54}

\author{David R. Ciardi}
\affil{Michelson Science Center/Caltech\\
 770 South Wilson Avenue, M/S 100-22\\
Pasadena, CA 91125}
\email{ciardi@ipac.caltech.edu}

\author{Cynthia G\'omez Mart\'in}
\affil{University of Florida, Department of Astronomy\\
211 Bryant Space Sciences Building, Gainesville, FL 32611}
\email{gomez@astro.ufl.edu}




\begin{abstract}

We present mid-infrared (10.4 \micron, 11.7 \micron, and 18.3 \micron) imaging
intended to locate and characterize the suspected protostellar components
within the Bok globule CB54. We detect and confirm the protostellar status for
the near-infrared source CB54YC1-II. The mid-infrared luminosity for
CB54YC1-II was found to be $L_{midir} \approx 8\ L_\sun$, and we estimate a
central source mass of $M_* \approx 0.8\ M_\sun$ (for a mass accretion rate of
${\dot M}=10^{-6}\ M_\sun\ yr^{-1}$). CB54 harbors another near-infrared
source (CB54YC1-I), which was not detected by our observations. The
non-detection is consistent with CB54YC1-I being a highly extinguished
embedded young A or B star or a background G or F giant. An alternative
explanation for CB54YC1-I is that the source is an embedded protostar viewed
at an extremely high inclination angle, and the near-infrared detections are
not of the central protostar, but of light scattered by the accretion disk
into our line of sight. In addition, we have discovered three new mid-infrared
sources, which are spatially coincident with the previously known dense core
in CB54. The source temperatures ($\sim100$K) and association of the
mid-infrared sources with the dense core suggests that these mid-infrared
objects may be embedded class 0 protostars.

\end{abstract}


\section{Introduction \label{intro-sec}}
Protostars are young stellar objects that are still in the process of
accreting the bulk of their material.  Class 0 sources have been proposed as
the evolutionary precursors to the class I protostars \citep{awb93, am94}.
Work by \citet{bontemps96} and \citet{saraceno96} suggested a direct
evolutionary sequence from the class 0 stage to the class I stage. However,
\citet{jhc01} have argued that class 0 protostars are located preferentially
in higher density regions and that class I young stellar objects are located
preferentially in lower density regions and, thus, may be at a comparable
evolutionary states.  To further complicate the distinction between class 0
and class I objects, some class I objects, which are viewed at high
inclination, may appear ``class 0-like'' because of the high optical depth
associated with viewing the disk (nearly) edge-on \citep{mi00}.

An example of this confusion may be seen in the binary system L1448N(A,B),
where one component of the system has the spectral energy distribution (SED)
of a class 0 protostar while the other has an SED of a class I protostar
\citep{crd03,olinger06}. This spread in apparent evolutionary status is also
seen in larger clusters. In Perseus, \citet{rebull07} found that sources
within clusters exhibited SEDs for a wide range in circumstellar environments,
suggesting class 0 to class II protostars within the same aggregate. Either
viewing geometry plays a significant role in our interpretation for each of
the SEDs, the aggregates span a significant age spread, or if the sources are
coeval, the disks/envelopes evolve faster than anticipated.

To help address these issues, it would be beneficial to study a set of young
stellar objects, located within the same environment, but isolated and free
from the influence of other active star formation. The Bok Globule CB54, known
to be a site of active star formation, may provide such an environment of
isolated aggregate star formation.

CB54 \citep[LBN 1042; ][]{cb88} is a $\sim100\ M_\odot$ Bok globule associated
with the Vela OB1 molecular cloud \citep[d$\approx$1500 pc; ][]{lh97,lwth97}.
Bok globules are small \citep[$10-100$ M$_\odot$;][]{cyh91}, isolated
molecular clouds, most of which have been identified via opaque patches in
optical images \citep{cb88,bhr95}.  Globules have been found to be sites of
star formation \citep[e.g.,][]{yc90, yc94a, yc94b, ay95, my95}, of both single
low-mass stars and multiple or binary stars \citep[e.g.,][]{yun96}.

\begin{figure}

    \includegraphics[angle=90,scale=0.7,keepaspectratio=true]{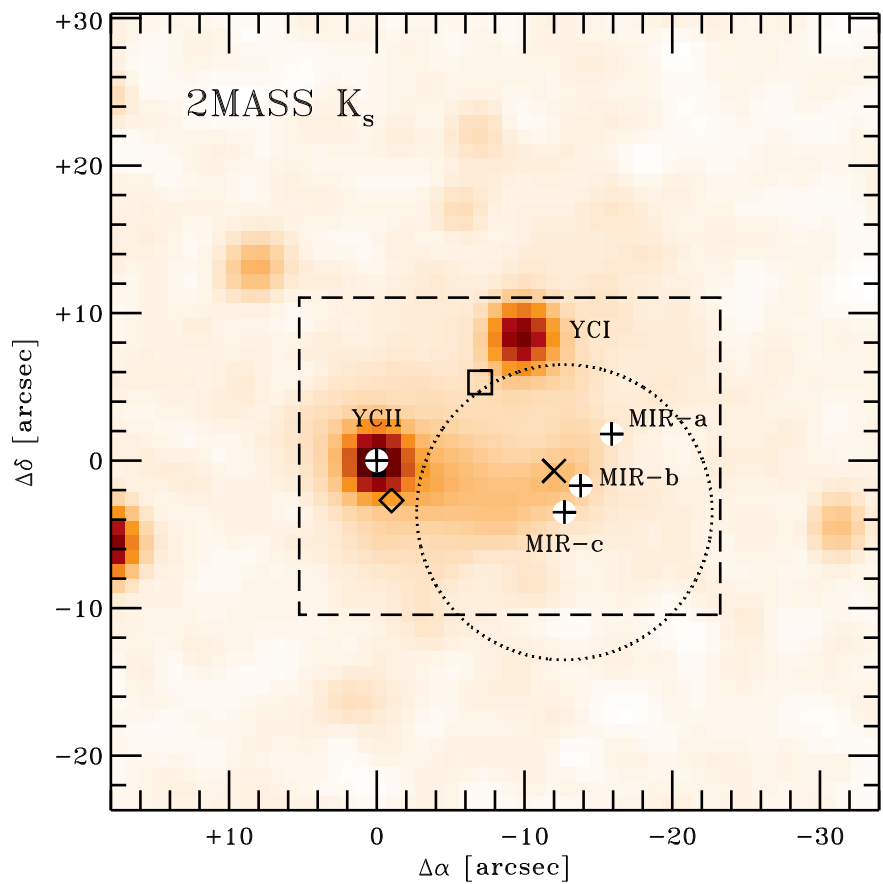}

    \figcaption{2MASS K$_S$ image of CB 54.  The dashed line delineates the
    area imaged with T-ReCS.  The near-infrared sources CB54YC1-I and CB54YC1-II
    are annotated, and the positions of the detected mid-infrared sources are
    marked with the filled white circles (see Figure \ref{mirimg-fig}).  The
    image (0,0) position is centered on CB54YC1-II. The circle (dotted line) is
    centered on the peak of 850 \micron\ core; the size of the circle
    represents the approximate size of the 850 \micron\ core
    \citep{henning01}. The cross marks the position of the IRAS source PSC
    07020-1618. The open diamond marks the position of the 8 \micron\ MSX
    source G228.9946-04.6200, and the open square marks the position for the cm
    source discovered with the VLA \citep{yun96b}. The image pixel scale is
    1$\arcsec\ {\rm pix}^{-1}$.\label{2mass-fig}}

\end{figure}

Active star formation in CB54 was first identified by the association of a
dense core with the IRAS point source PSC 07020-1618, which is located at the
center of a collimated molecular outflow \citep{yc94b}. Near-infrared imaging
revealed that CB54 actually contains two bright near-infrared sources
(CB54YC1-I, CB54YC1-II) and diffuse nebulosity of shocked H$_2$ emission
connecting the two sources \citep[see Figure \ref{2mass-fig} and
][]{yun96,khan03}.  The positions of CB54YC1-I and -II are offset from the
position of the IRAS point source (see Figure \ref{2mass-fig}), although the
IRAS beam size and PSC positional errors $(20\arcsec \times 4\arcsec)$ do make
it difficult to associate any one source with IRAS 07020-1618.

Based upon its near-infrared colors ($J-K = 5.29$ mag, $H-K = 2.58$ mag),
CB54YC1-II was classified as a class I young stellar object \citep{yun96}.
There is an MSX point source (G228.9946-04.6200), detected only in Band-A (8.3
\micron), within a few arcsecs of CB54YC1-II (see Figure \ref{2mass-fig}).
CB54YC1-I ($J-K = 4.34$ mag, $H-K = 1.63$ mag) was also classified as a class
I protostar \citep{yun96}, but despite its similar near-infrared brightness to
CB54YC1-II, it was not detected by MSX.  Unlike CB54YC1-II, the near-infrared
colors of CB54YC1-I could be explained with a highly extinguished ($A_V \sim
20$ mag) ``bare'' photosphere. VLA observations detected a 3.6 cm and 6 cm
source within a few arcseconds of CB54YC1-I (see Figure \ref{2mass-fig}),
possibly indicating a stellar wind or accretion shock \citep{yun96b,
moreira97}.

The position of the IRAS point source is spatially coincident with a dense
core revealed in sub-mm, mm, and molecular line mapping; the position of which
is offset from the positions of CB54YC1-I and CB54YC1-II but coincident with
the IRAS point source \citep[see Figure \ref{2mass-fig} and e.g.,
][]{wang95,zew96,henning01}. Molecular line observations also indicated the
presence of gravitational collapse in the core of CB54 \citep{wang95, ayc98}.
Water maser emission, almost exclusively associated with class 0 protostars in
regions of low-mass star formation \citep[e.g.,][]{furuya01}, was also
discovered in CB54 \citep{degm06,gomez06}.  All of this suggests that the star
formation in CB54 may be more substantial than revealed by the near-infrared
imaging alone.

We have observed the mid-infrared emission from the Bok globule CB54 at high
spatial resolution ($\sim 0\farcs5$) to clarify the evolutionary status of
CB54YC1-I and CB54YC1-II and to search for additional protostars embedded in
the globule core. Our work confirms the protostellar status of CB54YC1-II, but
indicates that CB54YC1-I may be a more evolved young stellar object or a
background giant star. In addition, we have discovered three new mid-infrared
sources which are spatially coincident with the dense core and may be class 0
protostars.

\section{Observations and Data Reduction\label{obs-sec}}

Mid-infrared imaging observations of CB54 were made on 2004 February 01 (UT)
using the Thermal Region Camera and Spectrograph \citep[T-ReCS; ][]{telesco98}
on the Gemini South 8 m telescope. T-ReCS utilizes a $320 \times 240$ pixel
Si:As blocked impurity band detector, with a spatial scale of $0\farcs089$
pixel$^{-1}$ and a field of view of $28\farcs8 \times 21\farcs6$.  The
observations were centered on the J2000 coordinates $(\alpha, \delta) =
(\alpha=07h04m21s,\ \delta=-16^{\circ}23\arcmin19\arcsec)$. Imaging was
obtained in three filters (N, Si-11.7 and Qa-18.3).  The on-sky alignment of
the T-ReCS field-of-view was chosen to cover the entire near-infrared
nebulosity, the two known near-infrared sources, and the peak of the
sub-millimeter (850 \micron) core (see Figure \ref{2mass-fig}).

A standard off-chip 15\arcsec\ north-south chop-nod sequence was employed with
total on-source integration times of 300 s per image.  Three exposures in the
Qa-18.3 filter were acquired for a total on-source integration time of 900 s.
Flux calibration was obtained from imaging of the standard star HD~32887 (see
the Gemini webpage for a compilation of mid-infrared standard stars and flux
densities.)\footnote{
http://www.gemini.edu/sciops/instruments/mir/MIRPhotStandards.html}. The
weather quality was listed as the 50th-percentile, and the seeing at 11.7
\micron\ was $\approx 0\farcs4$. A summary of the filters, frame times, total
integration time per filter, and associated air masses is given in Table
\ref{obssum-tab}.

\begin{figure}

    \includegraphics[angle=0,scale=0.7,keepaspectratio=true]{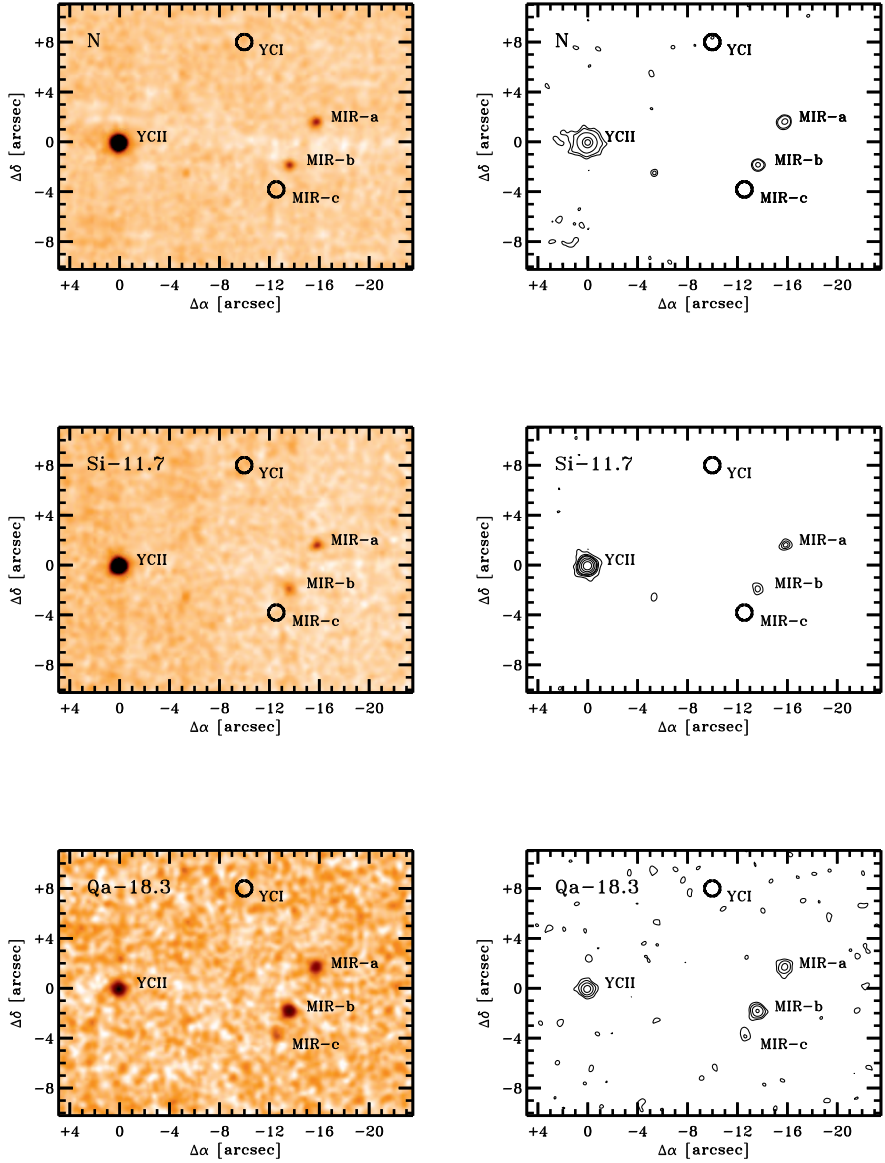}

    \figcaption{T-ReCS N-band, 11.7\micron\ and 18.3\micron\ images
    and contour plots of CB54. The (0,0) point of each image is
    centered on CB54YC1-II.  The images have been stretched by an
    inverse hyperbolic sine to enhance the contrast.  The detected
    mid-infrared sources are annotated. The position of CB54YC1-I
    (not detected by the mid-infrared observations) is marked in
    each image, and the position of MIR-c (detected only at
    18.3\micron) is marked in the N-band and 11.7\micron\ images.
    The N-band contour levels are $[0.03,0.05,0.15,1.5,3.5]$ mJy
    pix$^{-1}$; the 11.7 \micron\ contour levels are
    $[0.05,0.10,0.15,0.2,0.4,0.8,1.6]$ mJy pix$^{-1}$; the 18.3
    \micron\ contour levels are $[0.4,0.8,1.6,3.2]$ mJy pix$^{-1}$.
    The pixel scale for each image is 0$\farcs089\ {\rm
    pixel}^{-1}$. \label{mirimg-fig}}

\end{figure}

The data were reduced with custom-written IDL routines for the T-ReCS data
format. Four mid-infrared sources were detected by our observations, with no
evidence of extended or diffuse mid-infrared emission (see Figure
\ref{mirimg-fig}). Standard aperture photometry was performed using a
1\arcsec\ aperture radius.  Detection limits were tested by inserting fake
sources into the images and performing aperture photometry.  A summary of the
photometry (including estimated 1$\sigma$ upper limits) and relative
positional offsets with respect to CB54YC1-II is given in Table
\ref{photsum-tab}.

\section{Discussion \label{disc-sec}}

The near-infrared source CB54YC1-II is the brightest mid-infrared source and
is detected in each of the mid-infrared filters.  The other near-infrared
source CB54YC1-I was not detected in any of the mid-infrared imaging. In
addition, three new mid-infrared sources have been detected (MIR-a, MIR-b, and
MIR-c). MIR-a and MIR-b were detected in each of three mid-infrared filters,
while MIR-c was detected only at 18.3 \micron.  In the following sections, we
evaluate the properties of these sources and discuss the possible star
formation history of the globule.

\subsection{CB54YC1-II \label{ycii-subsec}}

\begin{figure}

    \includegraphics[angle=90,scale=0.7,keepaspectratio=true]{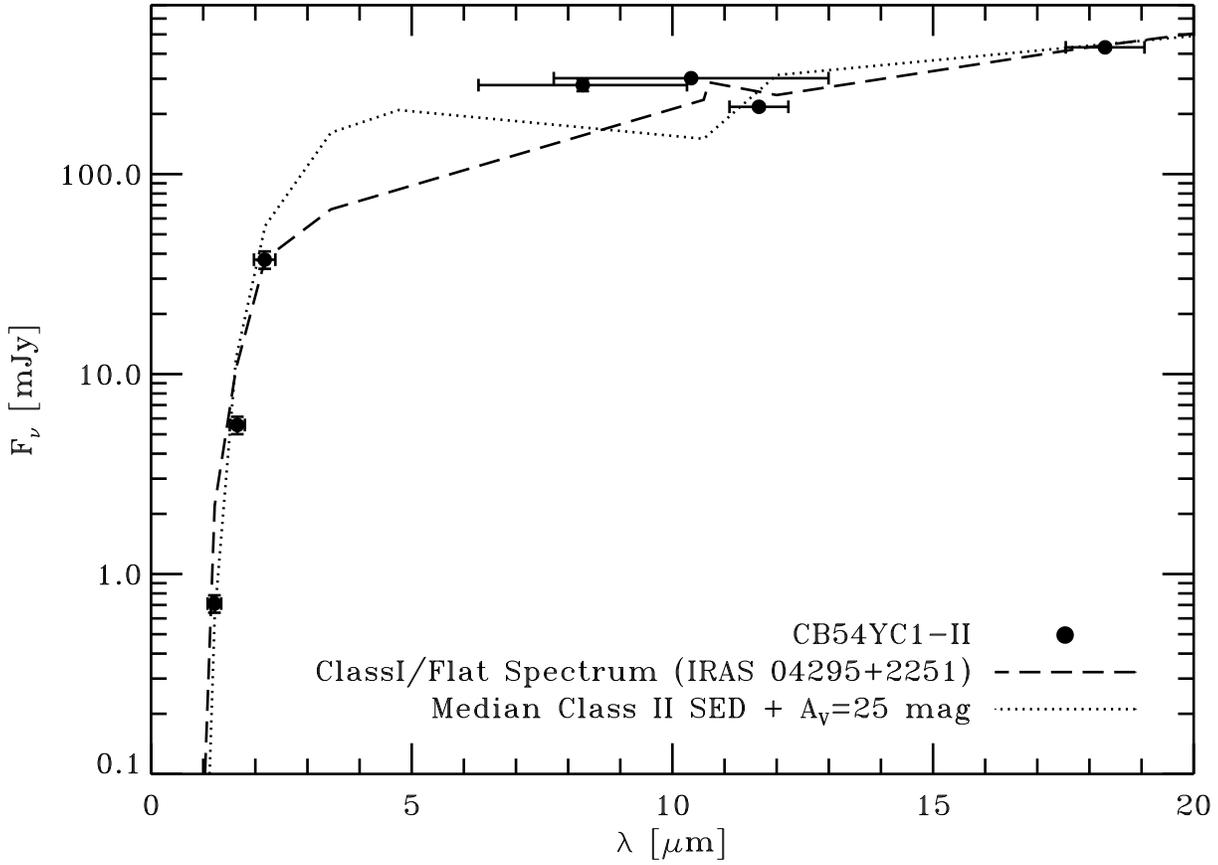}

    \figcaption{The SED of CB54YC1-II (solid points). The horizontal error
    bars represent the bandwidth of the filters.  The dashed line represents
    the SED for the class I protostar IRAS 04295+2251.  The dotted line
    represents a median TTS SED convolved with $A_V=25$ magnitudes of
    extinction. \label{yciised-fig}}

\end{figure}

CB54YC1-II, originally classified as a candidate class I protostar
\citep{yun96}, has a $2.2 - 10.3\ \micron$ spectral index $[ \alpha = - d\
{\rm log}(\nu F_\nu)/d\ \rm{log}(\nu)]$ of $\alpha = 0.34\pm0.01$, which is
consistent with the spectral index expected for a class I/flat spectrum young
stellar object. The SED of CB54YC1-II in Figure \ref{yciised-fig}.  For
comparison, the SED for a confirmed class I young stellar object \citep[IRAS
04195+2251;][]{eisner06}, with a similar spectral index $\alpha_{(2.2 - 10.6\
\micron)} \approx 0.3$, has been scaled to the SED of CB54YC1-II (see Figure
\ref{yciised-fig}), exhibiting good agreement between the SEDs.

If the mid-infrared emission is primarily the result of gravitational infall,
\citep[e.g.,][]{crd03}, the mid-infrared luminosity provides a means of
estimating the central protostellar mass. Integrating the SED from $1-20$
\micron, we estimate the mid-infrared luminosity to be $L_{midir} \approx
8\pm2\ L_\sun$, for an assumed distance of 1500 pc.  We estimate a central
source mass from the relation $L = (G\dot{M} M_*)/R_*$, where $\dot{M}$ is the
mass infall rate, $R_*$ is the source size, and $M_*$ is the source mass
\citep{sal87}. Using a standard $R_* = 3\ R_\sun$ protostellar radius
\citep{sst80} and typical mass accretion rates of $\dot{M} = 10^{-5} -
10^{-6}\ M_\sun\ {\rm yr}^{-1}$ \citep{kch93}, we estimate the central
protostellar mass for CB54YC1-II to be $M_* = 0.08 - 0.8\ M_\sun$.

A class II pre-main sequence star located behind a wall of extinction could
also explain the observed SED for CB54YC1-II.  In Figure \ref{yciised-fig}, a
median SED for T Tauri stars (TTS) \citep{dalessio99} has been scaled and
convolved with an extinction model \citep[R=3.1 assumed,][]{mathis90}.  In
order to match both the mid-infrared flux densities and the slope of the
near-infrared, the TTS SED must be extinguished by $A_V \approx 25$
magnitudes, and indeed, the position of CB54YC1-II in a JHK color-color ($J-H
= 2.71$ mag, $H-K = 2.58$ mag) diagram is consistent with a heavily
extinguished TTS \citep[see Figure 4 in ][]{haisch00}.

The average volume density of the CB54 envelope (i.e., not including the
central condensation which is offset from the near-infrared sources), as
derived from sub-mm (450 \& 850 \micron) imaging, is $\langle n_H \rangle
\approx 5 \times 10^4$ cm$^{-3}$ \citep{henning01}.  At 1500 pc, the projected
linear radius of the CB54 envelope is $r \approx 22500$ AU ($\approx
15\arcsec$). If we assume the globule is spherical, we can derive a peak
column density of $N_H \approx 4 \times 10^{22}$ cm$^{-2}$, which corresponds
to a visual extinction of $A_V \approx 20$ mag.  The extinction estimatation
does not take into account specific structure within the cloud including any
dense envelope which may immediately surround the source, but does indicate
the above derived extinction levels for CB54YC1-II are possible. Without a
more complete SED or spectroscopy, especially at $3 - 8$ \micron, it is
difficult to distinguish between the class I and class II models for
CB54YC1-II

\subsection{CB54YC1-I \label{yci-subsec}}

\begin{figure}

    \includegraphics[angle=90,scale=0.7,keepaspectratio=true]{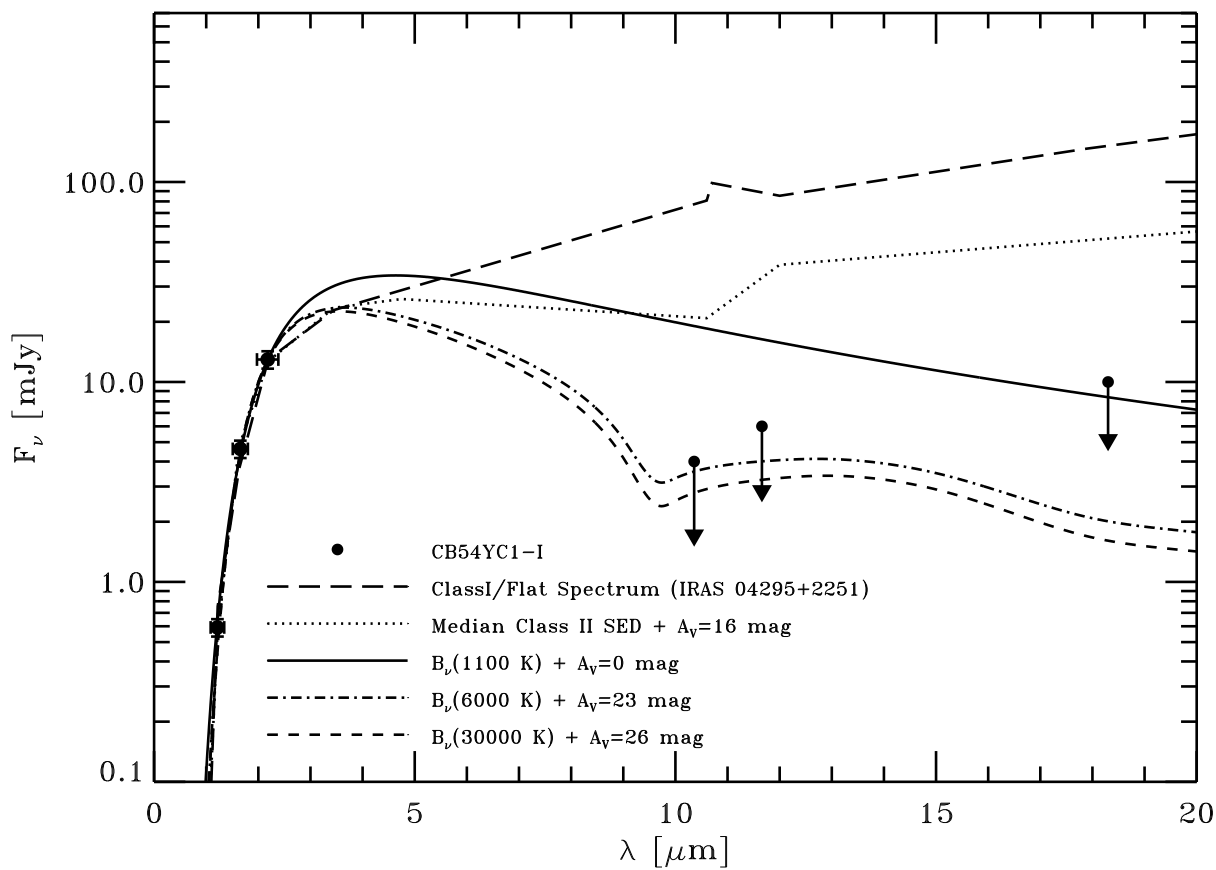}

    \figcaption{The SED of CB54YC1-I (solid points). Upper limits for the
    mid-infrared observations are represented by the downarrows.  The
    long-dashed line represents the SED for the class I protostar IRAS
    04295+2251.  The dotted line represents a median TTS SED convolved with $A_V=16$
    magnitudes of extinction.  The solid line is a 1100 K blackbody; the
    dash-dot line is a 6000 K blackbody convolved with $A_V=23$ magnitudes of
    extinction, and the short-dashed line is a 30000 K blackbody convolved
    with $A_V=26$ magnitudes of extinction.\label{ycised-fig}}

\end{figure}

CB54YC1-I was not detected in any of the three mid-infrared images, calling
into question the original class I protostellar classification. Scaling the
SED for the class I protostar (IRAS 04295+2251) to the JHK SED of CB54YC1-I,
the predicted mid-infrared flux densities for CB54YC1-I are $F_\nu \sim
80-100$ mJy at 11.7 \micron\ and $F_\nu \sim 150$ mJy at 18.3 \micron. The
predicted emission is $\sim 10\sigma$ above the detection limits (see Figure
\ref{ycised-fig}.

The JHK slope for CB54YC1-I is too steep to match the near-infrared SED for a
class II (TTS) pre-main sequence star. However, if TTS SED is modified with a
screen of foreground extinction, the near-infrared SED for CB54YC1-I can be
reproduced with a class II pre-main sequence model. In Figure
\ref{ycised-fig}, as was done for CB54YC1-II, the median TTS SED
\citep{dalessio99} has been scaled and convolved with an extinction model and
fitted to the JHK data for CB54YC1-I. For a best-fit extinction of $A_V = 15 -
17$ mag, the TTS SED can reproduce the near-infrared data.  However, the model
predicts mid-infrared densities ($F_\nu \sim 20-40$ mJy at 11.7 \micron\ and
$F_\nu \sim 50-70$ mJy at 18.3 \micron).  The predicted mid-infrared emission
for the TTS SED is $\sim 7\sigma$ above the detection limits.

It is possible that CB54YC-I is a more evolved star embedded in the globule or
simply a star background to the globule. To explore these possibilities, we
have fitted the JHK photometry with a blackbody function modified by a
line-of-sight extinction curve: $S_\nu = \Omega B_\nu(T)\exp{( {- {\rm
A}_\nu}/1.086 ) }$, where $B_\nu(T)$ is the Planck function, ${\rm A}_\nu$ is
the frequency-dependent extinction, and $\Omega$ is the solid angle.  At each
extinction value in the range from $A_V = 0 - 30$ mag ($\Delta A_V=0.1$ mag),
a range of temperatures ($T = 500 - 50000$ K in steps of 100 K) were tested.

\begin{figure}
    \vspace*{-0.5in}
    \includegraphics[angle=0,scale=0.7,keepaspectratio=true]{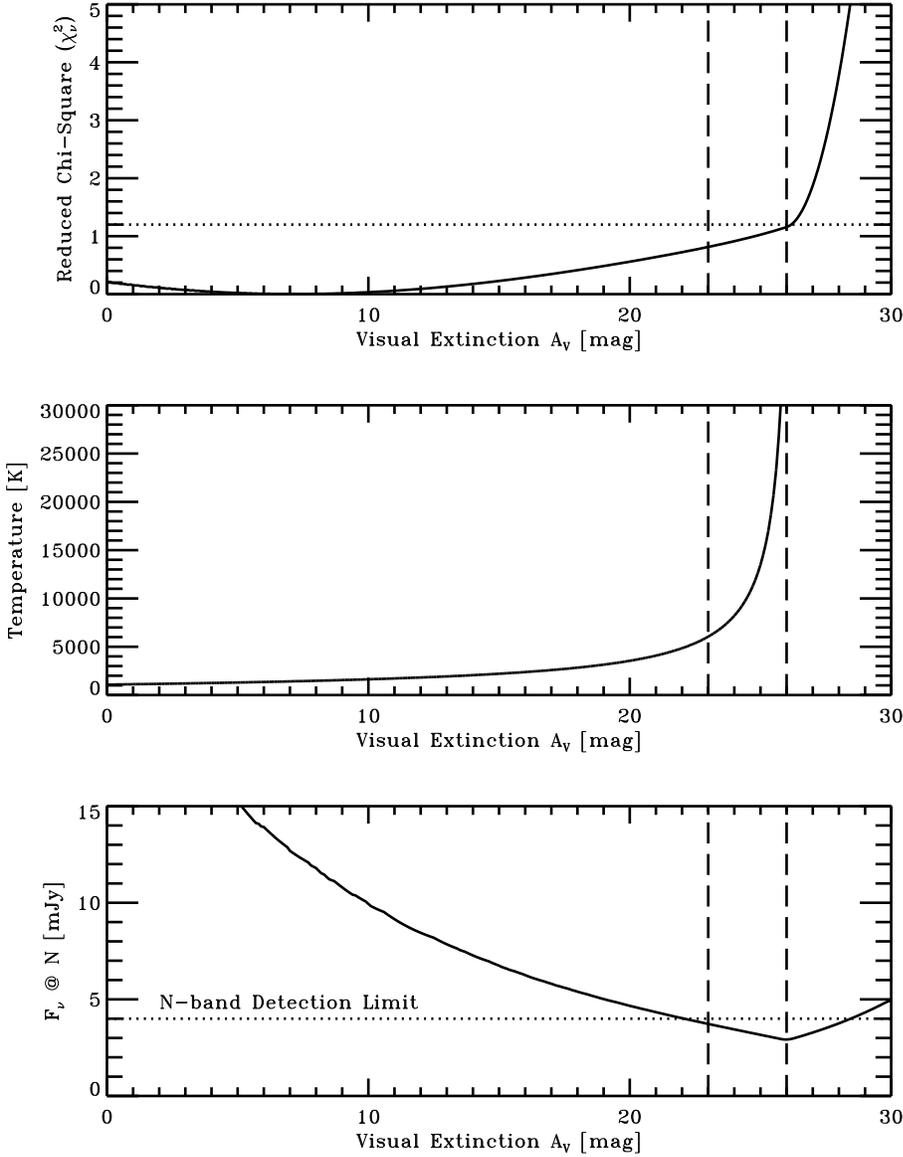}

    \figcaption{Reduced chi-squares ({\em top}), best-fit temperatures ({\em
    middle}), and predicted N-band flux densities ({\em bottom}), are plotted
    as a function of visual extinction for the blackbody+extinction models
    fitted to the JHK flux densities of CB54YC1-I.  The horizontal dashed line
    in the chi-square plot marks the sharp knee in the chi-square curve at
    $A_V=26$ mag ($\chi^2_\nu=1.2$). The horizontal dashed line in the flux
    density plot marks the 1$\sigma$ detection limit (4 mJy) of the N-band
    observations.  The vertical dashed lines delineate the extinction range of
    $23 \leq A_V \leq 26$ mag (see text for details).\label{avtemp-fig}}

\end{figure}

For a given extinction value, there is a unique blackbody temperature for
which the chi-square is a minimum, but there is no global minimum representing
a best fit the JHK data. The average temperature uncertainty for a given
extinction value is $\sim 500$ K. In Figure \ref{avtemp-fig}, the resulting
reduced chi-squares and temperatures for each of the trial extinctions are
plotted. The fitting uncertainty in the temperature for a given extinction
value is approximately 10\%.  The reduced chi-square curve is relatively flat
between $0 \leq A_V \leq 26$ mag.   Beyond $A_V = 26$ mag, the reduced
chi-square climbs above $\chi^2_\nu \approx 1$ and begins to diverge rapidly.
The best fit temperature at this boundary is $T\approx 30000$ K. Because of
the rapid change in the chi-square beyond this point, we regard this as the
upper bound for the extinction and source temperature of CB54YC1-I.

The lower bound to the extinction and temperature is constrained only by the
detection limits of the mid-infrared observations.  The combination of
temperature and extinction must be such that CB54YC1-I is not detected in all
three mid-infrared filters (N, Si-11.7, \& Qa-18.3).  For example, at zero
extinction ($A_V =0$ mag), the best fit temperature is $T \approx 1100$ K and
the reduced chi-square is $\chi^2_\nu < 1$, but the predicted mid-infrared
flux densities violate the non-detections at both N and 11.7 \micron\ (see
Figure \ref{ycised-fig}).

The blackbody plus extinction models only predict mid-infrared flux densities
below the detection limits in all three mid-infrared filters if $A_V \gtrsim
23$ mag.  As an example, the predicted N-band flux densities (the most
sensitive of the three observations) are plotted as a function of visual
extinction in Figure \ref{avtemp-fig} ({\em bottom}), showing that the
predicted flux density drops below the detection limit at $A_V \gtrsim 23$
mag. We, therefore, regard $A_V \approx 23$ mag and the corresponding
temperature ($T \approx 6000$ K) as the lower bound for CB54YC1-I. The lower
($A_V=23$ mag, $T=6000$ K) and upper ($A_V=26$ mag, $T=30000$ K) bounds for
the blackbody model fits to CB54YC1-I are shown in Figures \ref{ycised-fig}.

If the extinction is $A_V = 23$ mag, the temperature ($T \approx 6000$ K)
suggests that the CB54YC1-I could be an early-G or late-F star. A main
sequence dwarf of this spectral type has an absolute K magnitude of $M_K
\approx 2.7$ mag. With a measured K magnitude of $K=11.76$ \citep{yun96}, the
implied distance is only $\sim250$ pc, much too close to be extinguished by
CB54. If, however, CB54YC1-I is a G or F giant star, the star would be
approximately 4.0 magnitudes brighter and located at a distance of $\approx
1500$ pc, sufficiently distant to place CB54YC1-I behind the globule.

At the upper limit to the model fitting, ($A_V \approx 26$ mag, $T\approx
30000$ K), the temperature corresponds to a B0 star.  With an estimated
absolute magnitude of $M_K = -3$. CB54YC1-I would be at a distance of 2500 pc
which is far enough to be behind the globule.  The rarity and short lifespan
(10 MYr) of B0 stars and the required chance alignment with the globule seem
to make this a remote possibility.

If, instead, the extinction is near the middle of the extinction range ($A_V =
24 -25$ mag), the best fit temperature ($T \approx 10000-15000$ K) implies
that CB54YC1-I may be a young A or B star.  With an absolute K magnitude of
$M_K \approx -1.5 - 0$, this yields a distance of only $1200-1500$ pc placing
CB54YC1-I at a distance consistent with being an embedded young star.

An alternative explanation for CB54YC1-I is that the source is an embedded
protostar viewed at an extremely high inclination angle, and the near-infrared
detections are not of the central protostar, but of light scattered by the
accretion disk into our line of sight. Unfortunately, near-infrared photometry
alone can not distinguish these models.

\subsection{Mid-Infrared Sources \label{mir-subsec}}

\begin{figure}

    \includegraphics[angle=90,scale=0.7,keepaspectratio=true]{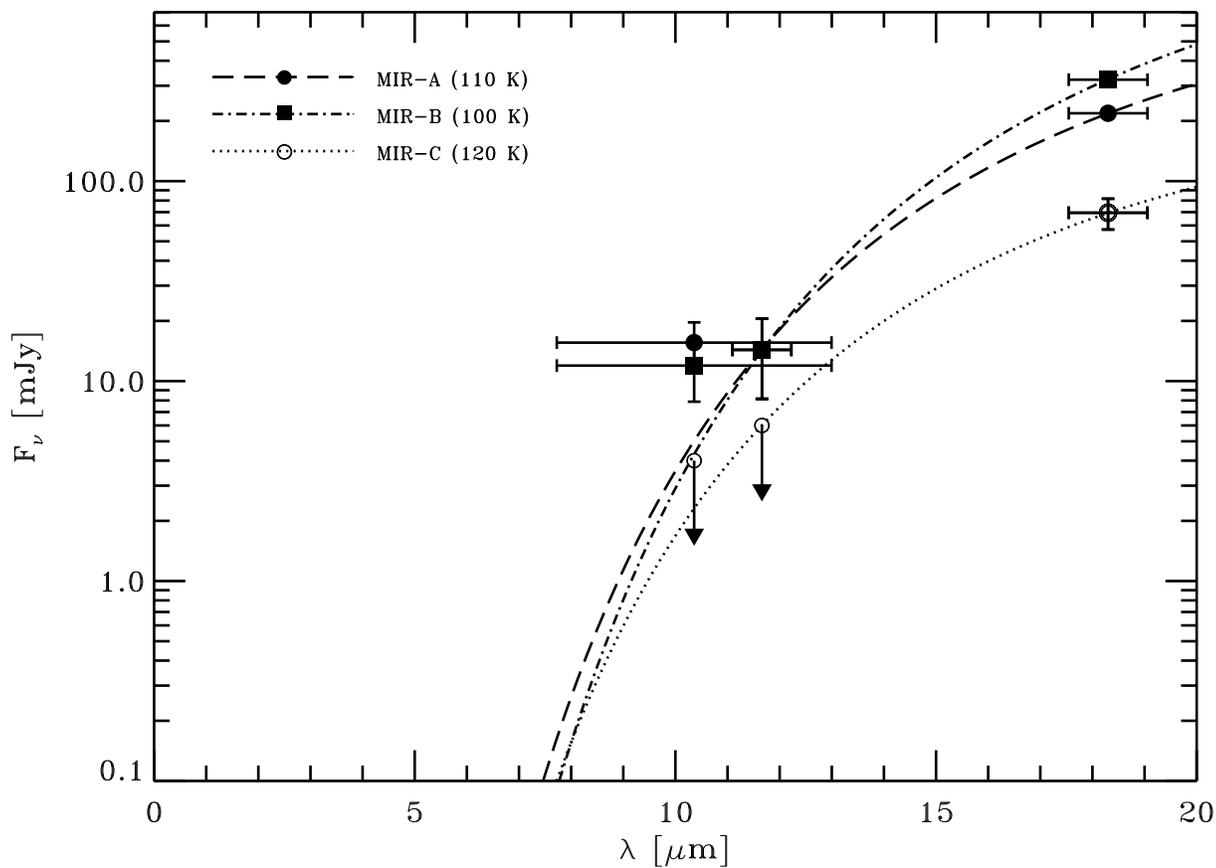}

    \figcaption{The SEDs of the mid-infrared sources
    MIR-a (solid squares), MIR-b (solid circles), and MIR-c (open circles).
    The 110 K and 100 K best-fit blackbody curves are shown for MIR-a (dashed)
    and MIR-b (dot-dash).  For MIR-c, the dotted line represents a 120 K
    blackbody which is fit to the 18.3\micron\ flux density and the
    11.7\micron\ upper limit. The horizontal error bars represent the
    bandwidth of the filters. \label{mirsed-fig} }

\end{figure}

Three new sources have been detected by our mid-infrared observations.  These
sources have no near-infrared counterparts.  The mid-infrared sources lie just
beyond the edge of the shocked H$_2$ emission and do not correspond to any of
the knots or condensations visible in the near-infrared diffuse emission
\citep[see Figure \ref{2mass-fig} and][]{yun96,khan03}. The mid-infrared
sources, however, are located within the boundaries of the dense core in CB54
and clustered near the position of the IRAS point source (see Figure 1).  The
two brightest sources (MIR-a, MIR-b) were detected in all three filters, but
MIR-c was detected only at 18.3 \micron.  A summary of the photometry is given
in Table \ref{photsum-tab}, and the SEDs for these sources are presented in
Figure \ref{mirsed-fig}.

To characterize and understand the relative temperatures of the mid-infrared
sources, a single temperature blackbody was fit to both MIR-a and MIR-b.  The
blackbody fits did not include the broad-band N (10.3 \micron) flux density
which are potentially contaminated with an unknown amount of amorphous
silicate, and were fit to the narrow-band 11.7 \micron\ and 18.3 \micron\ flux
densities. The best-fit blackbody (Figure \ref{mirsed-fig}) temperatures for
the two sources are quite similar ($T_A = 110 \pm 10$ K and $T_B = 100 \pm 10$
K) and are near what is expected for the bolometric temperatures of class 0
protostars \citep{awtb00}.  If the N-band photoometry is included in the fits,
the resulting temperatures increase by $10-20$ K.

Blackbody fits to the sub-mm and mm emission from the dense core yields a much
colder envelope temperature of 25 K \citep{lwth97}.  The summed flux densities
predicted by the mid-infrared blackbody fits is not sufficient to explain the
100 \micron\ flux density for IRAS PSC 07020-1618 (F$_\nu \approx 100$ Jy),
indicating that these class 0 protostars are harbored within the cold, dense
envelope.

If we assume that the mid-infrared emission is optically thin and little
emission is contributed from the cold envelope, we can estimate the
protostellar core mass associated with mid-infrared emission via
\begin{equation}
M_d = \frac{(16/3)\pi a \rho D^2}{Q_\nu B_\nu(T_d)}F_\nu
\end{equation}
where $F_\nu$ is the observed flux density at frequency $\nu$, $Q_\nu$ is the
grain emissivity at frequency $\nu$, $a$ is the grain radius, $\rho$ is the
grain mass density, $D$ is the distance to CB54, and $B_\nu$ is the Planck
function at dust temperature $T_d$. Assuming $a=0.5$ \micron, $\rho = 1$ g
cm$^{-3}$, $Q_\nu = 0.1(\lambda/\micron)^{-\alpha}$, and $\alpha=0.45$
\citep[e.g.,][]{mkv06}, we estimate dust masses for MIR-a and MIR-b of $M_a
\approx 0.014\ M_\odot$ and $M_b \approx 0.043\ M_\odot$, respectively. For an
average gas-to-dust mass ratio of 100, the central mid-infrared cores have
masses of $M_a \approx 1.4\ M_\odot$ and $M_b \approx 4.3\ M_\odot$.

MIR-c was detected only at 18.3 \micron, but if use the 11.7\micron\ uppper
limit to restrict the blackbody fitting, we find an upper limit to the
temperature of 120 K.  (see Figure \ref{mirsed-fig}). Coupled with the 18.3
\micron\ flux density, we estimate a limit to the total mass (gas+dust) of
$M_c \gtrsim 0.2\ M_\odot$.

\section{Conclusions \label{sum-sec}}

We have obtained high angular resolution $10-18$ \micron\ imaging of CB54, a
$\sim 100$ M$_\odot$ Bok globule known to harbor a dense core and two
near-infrared sources previously classified as class I young stellar objects.
We have detected only one (CB54YC1-II) of the two near-infrared sources,
confirming its protostellar evolutionary status.  Based upon the mid-infrared
luminosity, we estimate that the central protostellar mass for CB54YC1-II is
$M_* = 0.08 - 0.8\ M_\sun$, depending on the mass transfer rate.  The SED is
also consistent with a more evolved T Tauri star behind a screen of
extinction. Without a more complete SED, it is not possible to distinguish
between these models.

The other near-infrared source (CB54YC1-I) should have been detected if it
were a class I protostar similar to that CB54YC1-II. We find that the
near-infrared SED is consistent with the SED for a more evolved star
extinguished by the globule itself. CB54YC1-I may be a background F- or
G-giant or may be an embedded young A- or B-star.  An alternative explanation
for CB54YC1-I is that the source is an embedded protostar viewed at an
extremely high inclination angle, and the near-infrared detections are not of
the central protostar, but of light scattered by the accretion disk into our
line of sight.  High spatial resolution near-infrared polarimetry and/or
mid-infrared spectroscopy could be used to ascertain the status CB54YC1-I. If
CB54YC1-I is an embedded, young A or B star, its mass may be on the order of
$\sim 2-5\ M_\sun$.

Additionally, we have discovered three new mid-infrared sources (MIR-a, MIR-b,
and MIR-c) which are spatially coincident with both the position of the
associated IRAS point source and the center of the dense core in CB54.  These
sources are characterized with a 100 K blackbody, consistent the expected
bolometric temperature of a class 0 protostar.  Based upon the mid-infrared
emission, we have estimated the masses for these sources to be $\sim 4\
M_\sun$, $\sim 1.5\ M_\sun$, and $\sim 0.2\ M_\sun$.

If CB54YC1-I is indeed an embedded A or B star, it is interesting to speculate
that CB54YC1-I may have formed first and induced star formation further in the
cloud, through the interaction of its outflow/winds with remainder of the
globule. The total mass estimated for the sources within CB54 is about $10-15$
M$_\sun$ or about $10-15$\% of the total cloud mass. Such a sequential process
of star formation occurring in the Bok globule CB54 may be similar to what is
observed in other globules \citep{hws00, codella06}.  Spectroscopy and a more
complete SED in the mid-infrared and far-infrared is needed to disentangle the
possible spectral types and evolutionary states for the sources in CB54.

\acknowledgments

These observations were carried out during payback time to C. Telesco for
development of T-ReCS. The authors would like to thank Charlie Telesco, Chris
Packham, and Margaret Moerchen for collecting these data.  Portions of this
work were supported by the California Institute of Technology under contract
with the National Aeronautics and Space Administration.  C.~G.~M. acknowledges
support from a University of Florida Graduate Minority Fellowship, a SEAGEP
Fellowship, and NSF grants AST97-3367 and AST 02-02976. C.~G.~M. would like to
thank Eric McKenzie and Ana Matkovic for comments and suggestions.  Based on
observations obtained at the Gemini Observatory, which is operated by the
Association of Universities for Research in Astronomy, Inc., under a
cooperative agreement with the NSF on behalf of the Gemini partnership: the
National Science Foundation (United States), the Particle Physics and
Astronomy Research Council (United Kingdom), the National Research Council
(Canada), CONICYT (Chile), the Australian Research Council (Australia), CNPq
(Brazil) and CONICET (Argentina).  This research has made use of the NASA/IPAC
Infrared Science Archive, which is operated by the Jet Propulsion Laboratory,
California Institute of Technology, under contract with the National
Aeronautics and Space Administration.

\newpage

\begin{deluxetable}{cccccc}
\tablecolumns{6}

\tablewidth{0in}

\tablecaption{Summary of Observations \label{obssum-tab}}
\tablehead{
& \colhead{$\lambda_c$} & \colhead{$\Delta\lambda$} &
\colhead{Frame Time} & \colhead{On-Source} & \\
\colhead{Filter} & \colhead{($\mu$m)} & \colhead{($\mu$m)} & \colhead{(ms)}
& \colhead{(seconds)} & \colhead{Air Mass}
 }

\startdata
N & 10.36 & 5.27 & 25.8 & 304 & 1.11\\
Si-11.66 & 11.7 & 1.13 & 25.8 & 304 & 1.84\\
Qa-18.30 & 18.3 & 1.51 & 25.8 & 912 & 1.18-1.34

\enddata
\end{deluxetable}

\begin{deluxetable}{lccccccccc}
\rotate \tablecolumns{10}

\tablewidth{0in}

\tablecaption{Source Positions and Flux Densities \label{photsum-tab} }
\tablehead{
& & \colhead{J} & \colhead{H} & \colhead{K} & \colhead{MSX-A} & \colhead{N} & \colhead{Si-11.7}
& \colhead{Qa-18.3} &\\
& & \colhead{1.22 \micron} & \colhead{1.65 \micron} & \colhead{2.18 \micron} & \colhead{8.28 \micron}
& \colhead{10.36 \micron}  & \colhead{11.66 \micron} & \colhead{18.3 \micron} &\\
\colhead{Source} &  \colhead{$(\Delta\alpha,\
\Delta\delta)$\tablenotemark{a}} & \colhead{(mJy)} & \colhead{(mJy)}
& \colhead{(mJy)} & \colhead{(mJy)} & \colhead{(mJy)} &
\colhead{(mJy)} & \colhead{(mJy)} & \colhead{References}
 }

\startdata

YCII & $(0.0,0.0)$ &$0.71\pm0.07$& $5.56\pm0.56$ & $37.3\pm3.7$ & $278\pm19$ & $302\pm5$ &
    $217\pm6$ &$431\pm11$ &1,2\\

YCI & $(-10.0,+8.3)$&$0.59\pm0.06$& $4.63\pm0.46$ & $12.9\pm1.3$ & \nodata & $<4$ & $<6$ &
    $<10$&1,2\\

MIR-a & $(-15.8,+1.7)$ & \nodata & \nodata & \nodata & \nodata & $16\pm4$ & $14\pm6$ &
    $219\pm11$ &2\\

MIR-b & $(-13.6,-1.8)$& \nodata & \nodata & \nodata & \nodata & $12\pm4$ & $14\pm6$ &
    $322\pm11$ &2\\

MIR-c &$(-12.6,-3.8)$&  \nodata & \nodata & \nodata & \nodata &$<4$ &
    $<6$ &$70\pm12$ & 2

\enddata
\tablenotetext{a}{Positional offsets in arcsec from the source CB54YC1-II
($\alpha=07h04m21.7s,\ \delta=-16^{\circ}23\arcmin19\arcsec$ (J2000)).}
\tablerefs{1.~\citet{yun96}; 2.~This Work}

\end{deluxetable}

\newpage

\end{document}